\input harvmac
\noblackbox
\input epsf


\newcount\figno
\figno=0
\def\fig#1#2#3{
\par\begingroup\parindent=0pt\leftskip=1cm\rightskip=1cm\parindent=0pt
\baselineskip=11pt

\global\advance\figno by 1
\midinsert
\epsfxsize=#3
\centerline{\epsfbox{#2}}
\vskip 12pt
\centerline{\vbox{{\bf Figure \the\figno:} #1}}\par
\endinsert\endgroup\par}
\def\figlabel#1{\xdef#1{\the\figno}}
\overfullrule=0pt


\def\underarrow#1{\vbox{\ialign{##\crcr$\hfil\displaystyle
 {#1}\hfil$\crcr\noalign{\kern1pt\nointerlineskip}$\longrightarrow$\crcr}}}
%
\def\tilde{\widetilde}
\def\bar{\overline}

%

\font\cmss=cmss10
\font\cmsss=cmss10 at 7pt
\def\rlx{\relax\leavevmode}
\def\inbar{\vrule height1.5ex width.4pt depth0pt}
\def\IC{\relax\,\hbox{$\inbar\kern-.3em{\rm C}$}}
\def\IN{\relax{\rm I\kern-.18em N}}
\def\IP{\relax{\rm I\kern-.18em P}}
\def\IR{\relax{\rm I\kern-.18em R}}
\def\IC{{\relax\hbox{$\inbar\kern-.3em{\rm C}$}}}
\def\IZ{\relax\ifmmode\mathchoice
{\hbox{\cmss Z\kern-.4em Z}}{\hbox{\cmss Z\kern-.4em Z}}
{\lower.9pt\hbox{\cmsss Z\kern-.4em Z}}
{\lower1.2pt\hbox{\cmsss Z\kern-.4em Z}}\else{\cmss Z\kern-.4em
Z}\fi}
\def\IH{\relax{\rm I\kern-.18em H}}
\def\ZZ{\rlx\leavevmode\ifmmode\mathchoice{\hbox{\cmss Z\kern-.4em Z}}
 {\hbox{\cmss Z\kern-.4em Z}}{\lower.9pt\hbox{\cmsss Z\kern-.36em Z}}
 {\lower1.2pt\hbox{\cmsss Z\kern-.36em Z}}\else{\cmss Z\kern-.4em
 Z}\fi}
\def\narrowplus{\kern -.04truein + \kern -.03truein}
\def\narrowminus{- \kern -.04truein}
\def\narrowminussub{\kern -.02truein - \kern -.01truein}

\def\frac#1#2{{#1\over #2}}

\def\CN{{\cal N}}

\def\BR{\IR}
\def\BZ{\IZ}

\def\IZ{\relax\ifmmode\mathchoice
{\hbox{\cmss Z\kern-.4em Z}}{\hbox{\cmss Z\kern-.4em Z}}
{\lower.9pt\hbox{\cmsss Z\kern-.4em Z}}
{\lower1.2pt\hbox{\cmsss Z\kern-.4em Z}}\else{\cmss Z\kern-.4em
Z}\fi}

\def\ra{\rightarrow}

\def\RP#1{{\BR {\rm P}^{#1}}}

%
%
\def\eqnn#1{\xdef #1{(\secsym\the\meqno)}\writedef{#1\leftbracket#1}%
\global\advance\meqno by1\wrlabeL#1}
\def\eqna#1{\xdef #1##1{\hbox{$(\secsym\the\meqno##1)$}}
\writedef{#1\numbersign1\leftbracket#1{\numbersign1}}%
\global\advance\meqno by1\wrlabeL{#1$\{\}$}}
\def\eqn#1#2{\xdef #1{(\secsym\the\meqno)}\writedef{#1\leftbracket#1}%
\global\advance\meqno by1$$#2\eqno#1\eqlabeL#1$$}


\nref\juan{J. Maldacena, ``The Large N Limit of Superconformal 
Field Theories and Supergravity'', hep-th/9711200, Adv. Theor. Math. 
Phys.2,231,1998}

\nref\edig{E. Witten, ``Anti-de-Sitter Space and Holography'',
hep-th/9802150, Adv. Theor. Math. Phys.2,253,1998; 
S.S. Gubser, I.R. Klebanov and A.M. Polyakov,
``Gauge Theory Correlators from Noncritical String Theory'',
hep-th/9802109, Phys. Lett.B 428,105,1998}

\nref\evashamit{S. Kachru and E. Silverstein, 
``4-D Conformal Theories and Strings on Orbifolds'', hep-th/9802183,
Phys. Rev. Lett. 80,4855,1998}

\nref\mbsj{M. Berkooz and S.-J. Rey, ``Nonsupersymmetric Stable 
Vacua of M-theory'', hep-th/9807200, JHEP 9901:014,1999}

\nref\igorandco{I.R. Klebanov and A.A. Tseytlin,
"D-Branes and Dual Gauge Theories in Type 0 String", hep-th/9811035;
I.R. Klebanov and A.A. Tseytlin,
"Asymptotic Freedom and Infrared Behavior in the Type 0 String Approach to
Gauge Theory", hep-th/9812089;
I.R. Klebanov and A.A. Tseytlin,
"A Nonsupersymmetric Large N CFT from Type 0 String Theory",
hep-th/9901101;
A.A. Tseytlin and K. Zarembo,
``Effective Potential in Non-Supersymmetric $SU(N)\times SU(N)$ Gauge 
Theory and Interactions of Type 0 D3-Brane'',
hep-th/9902095}

\nref\nik{N. Nekrasov and S.L. Shatashvili, 
``On Nonsupersymmetric CFT in Four Dimensions'', hep-th/9902110}

\nref\nifty{M. Bershadsky, Z. Kakushadze and C. Vafa,
``String Expansion as Large N Expansion of Gauge Theory'',
hep-th/9803076, Nucl. Phys. B523,59, 1998;
Z. Kakushadze,
``Gauge Theories from Orientifolds and Large N Limit''
hep-th/9803214, Nucl. Phys. B529, 157,1998;
M. Bershadsky and A. Johansen,
``Large N Limit of Orbifold Field Theories'', hep-th/9803249,
Nucl. Phys. B536,141, 1998}

\nref\bz{T. Banks and A. Zaks, "On the Phase Structure of Vector-Like 
Gauge Theories with Massless fermions'', Nucl. Phys. B196,189,1982.}

\nref\ja{J. Maldacena, J. Michelson and A. Strominger,
``Anti-de-Sitter Fragmentation'', hep-th/9812073}

\nref\mbakthree{M. Berkooz and A. Kapustin, work in progress}

\nref\edtor{E. Witten, ``Baryons and Branes in AdS'', 
hep-th/9804001, JHEP 9807:006,1998}

\nref\sethi{S. Sethi, ``A Relation between N=8 Gauge Theories in 
Three Dimensions'', hep-th/9809162,
JHEP 9811:003,1998}

\nref\mbakone{M. Berkooz and A. Kapustin, ``New IR Dualities in 
Supersymmetric Gauge Theories'', hep-th/9810257, To be published in JHEP.}
 
\nref\duff{M.J. Duff, B.E.W. Nilsson and C.N. Pope, Phys. Rep. 130,1,1986
; M.J. Duff, B.E.W. Nilsson and C.N. Pope, ``Spontaneous Supersymmetry
Breaking by the Squashed Seven-Sphere'', Phys. Rev. 50,2043,1983,
Erratum 51, 846,1983.}

\nref\itzhaki{N. Itzhaki, J. Maldacena, J. Sonnenschein and 
S. Yankielowicz, ``Supergravity and the Large N Limit of Theories with
Sixteen Supercharges'', hep-th/9802042, Phys. Rev. D58,46,1998}

\nref\ansz{A. Fayyazuddin and M. Spalinski,
``Large N Superconformal Gauge Theories and Supergravity Orientifolds'',
hep-th/9805096, Nucl. Phys. B535,219,1998; 
O. Aharony, A. Fayyazuddin and J. Maldacena, ``The Large N 
Limit of N=2,1 Field Theories from Threebranes in F-Theory'',
hep-th/9806159, JHEP 9807:013,1998;
A. Kehagias, ``New Type IIB Vacua and Their F-Theory Interpretation'',
hep-th/9805131, Phys. Lett. B435,337,1998} 

\nref\natisixteen{N. Seiberg, ``Notes on Theories with Sixteen 
Supercharges'',
hep-th/9705117, Nucl. Phys. Proc. Suppl.67:158-171,1998}

\nref\seven{B. Biran, A. Casher, F. Englert, M. Rooman and P. Spindel,
``The Fluctuating Seven Sphere in Eleven Dimensional Supergravity'',
Phys. Lett. 134B,179,1984;
L. Castellani, R. D'Auria, P. Fre, K.Pilch and P. Van Nieuwenhuizen,
``The Bosonic Mass Formula for Freund-Rubin Solution of d=11 Supergravity
General Coset Manifolds'',
Class. Quant. Grav. 1,229,1984}

\nref\sevenf{O. Aharony, Y. Oz and Z. Yin, 
``M-Theory on $AdS(P)\times S(11-P)$ and Superconformal Field Theories'',
hep-th/9803051, Phys.Lett.B430,87,1998}


\Title{\vbox{\hbox{hep-th/9903195}\hbox{IASSNS-HEP-99/31}}}
{\vbox{\centerline{A Comment on Nonsupersymmetric Fixed Points and}
\centerline{}
\centerline{Duality at large N}}}
\smallskip
\centerline{Micha Berkooz\footnote{$^1$} {berkooz@sns.ias.edu} and
Anton Kapustin\footnote{$^2$} {kapustin@sns.ias.edu} }
\vskip 0.12in
\medskip\centerline{\it School of Natural Sciences}
\centerline{\it Institute for Advanced Study}\centerline{\it
Princeton, NJ 08540, USA}

\vskip 1in

We review some of the problems associated with deriving field
theoretic results from nonsupersymmetric AdS, focusing on how to
control the behavior of the field theory along the flat directions. We
discuss an example in which the origin of the moduli space remains a
stable vacuum at finite $N$, and argue that it corresponds to an
interacting CFT in three dimensions. Associated to this fixed point is
a statement of nonsupersymmetric duality. Because $1/N$ corrections
may change the global picture of the RG flow, the statement of duality
is much weaker than in the supersymmetric case.

\vfill\eject
\vskip 0.1in

\newsec{Introduction}

The AdS/CFT correspondence \juan,\edig\ is a powerful tool for
studying the large $N$ limit of field theories. By now a significant
number of matches has been made between the dynamics of gauge theories
and the dynamics of supergravity in the corresponding backgrounds. For
the most part this analysis has been carried out in a supersymmetric
setting.

An interesting question is whether one can use gravity to understand
the dynamics of nonsupersymmetric conformal field theories at large
N. To answer this question one is led to study string theory/M-theory
backgrounds of the form $AdS_p\times M_q$ where $M_q$ is a compact
manifold which breaks supersymmetry (either via orbifolding a
supersymmetric manifold \evashamit, or by other means
\mbsj). Another approach (related to the previous one \nik) 
uses type 0 string theory \igorandco.

When discussing nonsupersymmetric theories one usually appeals to
classical 11D supergravity (i.e., the leading term in the momentum
expansion) or to classical string theory, both of which correspond to
$N=\infty$. In trying to extend the discussion to large but finite N
one generically runs into problems.  In \mbsj\ the following two
problems were discussed:

\item{1.} If for $N=\infty$ there are fields whose 
masses are at the Breitenlohner-Freedman unitarity bound, then these
masses might be pushed below the bound by $1/N$ corrections.

\item{2.} If there are massless fields (i.e. fields 
that correspond to marginal operators at $N=\infty$) which are
invariant under all the symmetries, then $1/N$ corrections may shift
their VEVs significantly and there may not be a stable vacuum for
finite $N$, or if such vacuum exists, it may be qualitatively
different from the $N=\infty$ starting point.

\noindent It was shown in \mbsj, however, that it is easy to construct
models in which these problems do not arise.

Another problem which we will discuss in this paper is that of the
fate of flat directions present at $N=\infty$. Many nonsupersymmetric
gauge theories converge, in some formal sense at least, to a theory
with sixteen supercharges as $N\ra\infty$ \nifty, so in this limit the
scalar potential has flat directions. These flat directions are
typically lifted by $1/N$ effects, as a result of which the fields are
either driven away from the origin or attracted to the origin (or a
combination of both in different directions). In the former case the
vacuum at the origin is destabilized (in fact, the theory may not have
any stable vacuum at all), while in the latter case the origin is at
least perturbatively stable.  In the latter case there is generically
a mass gap, explaining why it is so hard to construct
nonsupersymmetric CFTs when scalars are present (there are however
examples of nonsupersymmetric fixed points with fermions in the weak
coupling regime \bz).  In this paper we will discuss a 2+1 dimensional
example in which the flat directions are lifted in a way which drives
the fields to the origin, nevertheless the theory does not become
massive and trivial there.

Before proceeding it is worth mentioning some open problems. The main
open problem is that it is not clear whether the expansion around
$N=\infty$ is only formal, or whether it can be used to really
approximate the physics at finite $N$. In backgrounds that correspond
to weakly coupled string theory there is a genus expansion which is an
expansion in $1/N$. If the contribution of each genus is finite then
there is a valid $1/N$ expansion. However, models in the perturbative
stringy regime, for example those based on D3-branes, run into problem
2 (the dilaton is always a dangerous massless field). In the strong
coupling regime (M-theory or type IIB string theory near its self-dual
points) it is not clear whether quantum corrections are small. More on
this point will appear in \mbakthree.

Another open problem is the issue of nonperturbative instabilities
which describe tunneling in the bulk. Presumably these effects are 
exponentially small at large $N$. Not much is
known about such instabilities (see however \ja), and we will not
change this situation here.

\newsec{The Example}

The example that we will focus on is that of M-theory on\foot{the 
spectrum is related to that of $AdS_4\times S^7$. 
The spectrum of the latter is 
computed in \seven\ and compared to field theory expectations in
 \sevenf.} $AdS_4\times {\bf S}^7/\BZ_2$ . This background is obtained
 by probing different kinds of $\BR^8/\BZ_2$ orbifolds of M-theory
 with either M2-branes or anti-M2-branes.

The two kinds of $\BR^8/\BZ_2$ orbifolds differ by the charge of the
singularity. The first one, which we call the A-orbifold, has membrane
charge $-1/16$, while the other one, which we will call the
B-orbifold, has charge $3/16$~\sethi. Both orbifolds preserve sixteen
supercharges, the same supercharges as those preserved by an M2-brane
parallel to the orbifold plane.  Hence probing the orbifold
singularities by M2-branes yields $\CN=8$ field theories in three
dimensions.  Supersymmetry implies that when the charge of the orbifold
singularity is positive (relative to that of the M2-brane) the long range 
gravitational field of the singularity is as if it had a positive mass;
contrary-wise, if the charge is negative, then the mass is negative
(this, for example, can be deduced from the cancelation of forces
between the M2-brane and the singularity).

For both singularities the near-horizon geometry in the limit of large
number of probes $N$ is $AdS_4\times \RP7$. The only difference
between the two backgrounds is the torsion class \edtor\ in
$H^4(\RP7,\BZ)=\BZ_2$ which specifies how a membrane propagating in
this background is to be quantized \refs{\edtor,\sethi,\mbakone}. The
A-singularity corresponds to a trivial torsion class, while the
B-singularity corresponds to a nontrivial one. In the large $N$ limit
the curvature is small, and M-theory on $AdS_4\times \RP7$ reduces to
supergravity on the same background. Since supergravity is insensitive
to the torsion, the supergravity spectrum will be exactly the same for
the two backgrounds. In this limit, the difference in the torsion
class becomes visible only if one considers solitonic objects
(M2-branes and M5-branes) wrapping nontrivial cycles of $AdS_4\times
\RP7$.

Similarly we can probe the A and B singularities with
anti-M2-branes. This yields models without any supersymmetry. The near
horizon geometry in this case is the ``skew-whiffed'' $AdS_4\times
\RP7$
\duff. The usual logic of the AdS/CFT correspondence leads to the 
conclusion that M-theory on a ``skew-whiffed'' $AdS_4\times \RP7$
describes a nonsupersymmetric CFT on the boundary. The backgrounds
obtained from the A and B singularities differ only by a torsion class
which does not affect the Kaluza-Klein spectrum.

Both A and B singularities can be regarded as a strong-coupling
limit of certain orientifold backgrounds in IIA string
theory~\sethi,\mbakone.  An $O2^-$ plane lifts to an M-theory
background of the form $(\BR^7\times {\bf S}^1)/\BZ_2$ which has two
orbifold singularities of type A. An $O2^+$ plane lifts to the same
orbifold, except that one singularity is of type A, and the other one
is of type B. Finally, an $\tilde{O2}^+$ plane (which is an $O2^-$
plane with a half-D2-brane stuck to it) lifts to a pair of
B-singularities. These IIA backgrounds can be probed with (anti-)D2
branes, which lift to (anti-)M2-branes of M-theory. Thus the $\CN=8$
CFTs described by M-theory on $AdS_4\times
\RP7$ are related to $\CN=8$ gauge theories on
D2-branes, while the $\CN=0$ CFTs described by M-theory on the
``skew-whiffed'' $AdS_4\times \RP7$ are related to the gauge
theories on anti-D2-branes. The precise nature of this relation will be
discussed in section 4. In this paper we will focus on the $\CN=0$ case.

Reference \mbsj\ discusses some aspects of supergravity on the
``skew-whiffed'' $AdS_4\times \RP7$. It was shown there that the
Kaluza-Klein spectrum has neither massless charged scalars, nor modes
saturating the Breitenlohner-Freedman bound. As explained in the
introduction, this implies that the ``skew-whiffed'' $AdS_4\times
\RP7$ avoids some immediate problems of nosupersymmetric
compactifications.  In the next section we will address another
potential problem associated with the presence of flat directions at
infinite $N$. We will argue that for the B-singularity  the potential
generated along the flat directions at large but finite $N$ does not
change the vacuum significantly. The model corresponding to the
A-singularity  is apparently destabilized by $1/N$ corrections.

\newsec{Lifting of the flat direction}

We are therefore interested in discussing anti-M2-branes probing an
$A$ or $B$ $\BR^8/\BZ^2$ singularity. Equivalently one may consider
M2-branes probing the charge-conjugated singularities which we will
call ${\bar A}$ and ${\bar B}$. In this section we will use the latter
viewpoint.

At leading order in $N$ there are flat directions which correspond to
moving the branes away from the singularity and away from each other
\foot{We are referring to the flat directions of 
the fixed point theory in the IR rather than to those of the UV theory
which flows to it.}. This can be seen in several ways, but in general
one expects \nifty\ that at $N=\infty$ the structure of the flat
directions is the same as in the corresponding $\CN=8$ theory.
 
To obtain some information about the potential along the flat directions
one can do a long distance M-theory computation: one can place the branes
at a distance $r>> l_p$ from the singularity and determine, based on
the charge and mass of the singularity, whether there is an attractive
or repulsive force between the branes and the singularity. This
computation has little to do with field theory, since the branes
are in the asymptotically flat region. However, because this
computation depends on the mass and charge of the singularity in the
same way as the correct near horizon computation, it distinguishes
correctly between attractive and repulsive potential.

Using this approach one can also see that the potential is subleading
in $1/N$. The leading term in the long distance computation ($r>>l_p$) is
nominally of order $N^2$ (coming from all pairwise interactions
between the branes), but because this is the same as in the $\CN=8$
theory it is $N^2\times 0=0$. On the other hand, the interaction
between the singularity and the branes is of order $N$, because there
is only one singularity.

The computation that we would like to do is to check the stability of
the AdS to fragmentation along the flat directions in the near horizon
geometry. The idea is to separate the branes into several
clusters and compute the potential as a function of separation. 
For simplicity we will focus on the case of a single cluster away from the
singularity (i.e., two clusters which are the images of each other).

\subsec{The approximate solution along the flat directions}

We will start with the supergravity solution representing two clusters
of M2-branes in flat space and then orbifold this solution. The metric
for several parallel D3-branes in flat space was written in \juan\ and
it is straightforward to generalize the ansatz \ansz\ to M2-branes:

\eqn\fltdsln{ds^2 = f^{-2/3} dx^2 + f^{1/3}(dr^2+r^2d\Omega^2)}
$$G_{x^0x^1x^2r^i}\propto \partial_{r^i} f^{-1}(r),$$ where $G$ is the
4-form field strength and $f$ is an harmonic function of the 8-vector
$r$. To obtain the situation with two clusters each containing $N$
M2-branes we set
$$f(r)={Nl_p^6\over |r-a|^6} + {Nl_p^6\over |r+a|^6},$$ where the
8-vector $a$ is the position of the cluster. From the field theory
point of view it is convenient to do a rescaling $u^i=r^i/l_p^{3\over
2}$ \juan.

Next we want to orbifold this background. Orbifolding introduces an
$\BR^8/\BZ_2$ singularity at $r=0$. To facilitate the analysis of this
background it is convenient to further rescale the coordinates so that
the metric near the origin is the canonical flat metric on $\BR^{11}$:
\eqn\rscl{y^i= \biggl({2N l_p^6\over a^6}\biggr)^{-{1\over3}}x^i,\ 
          z^i= \biggl({2N l_p^6\over a^6}\biggr)^{1\over 6}r^i,} after
which the metric and the 4-form are given by the same ansatz but with
the following harmonic function: ${\hat f}$:
$${\hat f}=
{1/2 \over \bigl|n - {z\over (2N)^{1/6}l_p}\bigr|^6}  +
{1/2 \over \bigl|n + {z\over (2N)^{1/6}l_p}\bigr|^6},$$
where $n$ is a unit 8-vector in the direction of $a$.

Since the metric near the origin is the canonical one, and for large
$N$ all curvatures and field strengths are small there, it is easy to
insert the fields of the $\BZ_2$ singularity at $z=0$. One can
identify the following regions in the orbifolded background:

1. $z^2<l_p^2$: inside this region the curvature and the field
strength produced by the singularity are large. Our knowledge of this
this region is not better or worse than that of the $\BR^8/\BZ_2$
singularity in flat space. The fields due to the clusters of M2-branes
(the curvature and the 4-form) are of order $1/N^{1\over6}$ there.

2. The fields produced by the singularity and the fields produced by the
branes are comparable when $${1\over z^7}\sim {1\over N^{1\over 6}}.$$ 
At this point both are weak and can be treated using perturbation theory
around flat space (locally).

3. At $z\sim N^{1\over 6} n$ we approach the cluster of M2-branes
around which the space looks like $AdS_4\times {\bf S}^7$. This
describes an ${\CN=8}$ IR fixed point to which our theory flows along
this flat direction.

In the region $z>l_p$, the fields produced by the singularity are
small, and so are the fields of the original background. The gravity
background is therefore under control, and furthermore, the
corrections to the background due to the introduction of the
singularity are small as well. In the following subsection we will
extract the influence of this small correction on the potential along
the flat directions.

\subsec{The potential along the flat directions}

We would like to know whether, upon the introduction of the
singularity, there is a potential which drives the center of the
cluster to the origin or repels it. This potential is subleading in
the $1/N$ expansion and can be easily computed if one neglects the
back-reaction of the singularity on the rest of the geometry. We saw
above that this approximation is valid for $z>l_p$.

Within this approximation the computation is straightforward. If we
were allowed to choose the mass ($m$) and charge ($Q$) of the
singularity arbitrarily (the charge is measured relative to the charge
of the M2-branes), then there would be a line in the $Q-m$ plane,
$Q=m$ in appropriate units, on which supersymmetry is preserved. A and
B singularities correspond to two points on this line (A has negative
charge, while B has positive charge). The points corresponding to the
$\bar A$ and $\bar B$ singularities which break supersymmetry also
have charges of opposite sign and lie on on the line $Q=-m$. Clearly
the sign of the potential will change when going from one side of the
line $Q=m$ to the other. Hence one of the SUSY-breaking singularities
will attract the two clusters of branes, and the other will repel
them.

In more detail, the computation goes as the follows. When we take into
account the singularity the action is
\eqn\actb{{\cal L}= {\cal L}_0 + 
m\int_{r=0} d^3x \sqrt{g_{ind}} + Q\int_{r=0} C^{(3)}} where ${\cal L}_0$
is the usual action of 11D supergravity and $g_{ind}$ is the
determinant of the induced metric on the plane $r=0$. The fields in
${\cal L}_0$ are the same as in the supersymmetric case, except for a
two-fold identification due to orbifolding. The terms localized at
$r=0$ are due to the mass and charge of the singularity. To compute
the leading contribution to the potential in the no-back-reaction
approximation one has to insert the ansatz \fltdsln\ for the two
symmetrically separated clusters into this action.

The terms that we are interested in are the kinetic terms for
$a^i(x^\mu)$ (we allow $a$ to depend slowly on $x^\mu$) and the terms
that encode the interaction of clusters with the singularity. The
latter are proportional to $\int_{r=0} dx C^{(3)}$ (the gravitational
term gives an equal contribution as can be seen by comparison with the
supersymmetric case). This gives a term in the effective Lagrangian
for $a$ of the form
$${1\over N}\int d^3x (U^i)^6,$$ where $U$ is the field theory
quantity with dimension $1/2$ ($U^i=a^i/l_p^{3/2}$).

The kinetic term is also easy to evaluate. The functional dependence
is determined by spontaneously broken scale invariance to be
proportional to
$$\int d^3x (\partial_\mu U^i)^2.$$ The coefficient
in front of this term is of order $N$. This can be seen
by rescaling the coordinates $x$ so that the
entire metric in the new coordinates is proportional to
$N^{1\over3}$. In this setup it is easy to obtain the $N$-scaling of 
${\cal L}_0$ and therefore the $N$-scaling of the kinetic term.

The result of this computation is that for a singularity with negative 
charge ($\bar B$) there is an attractive potential along the flat
directions, while for $\bar A$ the potential is repulsive. Furthermore, since the potential is suppressed by powers of $N$, it is small at large $N$, and the no-back-reaction approximation is self-consistent.

\newsec{Nonsupersymmetric Duality}

\subsec{Weakness of nonsupersymmetric duality}

The statement that we are after is that of IR duality, i.e., we would
like to exhibit two distinct (weakly coupled) theories in the UV which
flow in the IR to the fixed point described above. However, the
duality that we obtain here will be considerably weaker than the one
obtained in cases with higher supersymmetry.

\medskip
\noindent{\it Field theory considerations}

The reason that the duality is weaker is the following. Let us first
consider the case $N=\infty$. In this case the theory is a projection
of the $\CN=8$ theory, in the sense that its dynamics is the same as
in the latter, except that we restrict our attention to a subset of
operators \nifty. The dynamics of the $\CN=8$ theory is well
understood \natisixteen\ and it is known that at the origin of its
moduli space it flows from a free UV fixed point to an interacting
superconformal IR fixed point.

Consider now the $1/N$ corrections to the RG flow. They are present
everywhere along the RG trajectory. Such corrections, even though they
are small at each point in the field theory parameter space, can
change the global picture of the RG flow. Therefore they may change
the statement that the theory flows from the gaussian fixed point in
the UV to the interacting IR.

Nevertheless, even with $1/N$ corrections taken into account, there
exists an RG trajectory which ends at the IR fixed point and passes at
a distance of order $1/N$ from the gaussian fixed point. Therefore, if
one wishes to ``land'' at the IR fixed point, one needs to fix a
cutoff and add, besides the relevant perturbation that already exists
in the $N=\infty$ theory, other operators with fine-tuned coefficients
suppressed by powers of $1/N$. In principle, at each order in $1/N$
expansion one will have to tune the coefficients of all operators
allowed by symmetries, including nonrenormalizable ones (Of
course, we do not need to tune these infinite number of coefficients
independently since there would be an entire submanifold of
trajectories which passes close to the gaussian UV and ends in the
interacting IR). Note that at large $N$ we are still close to the
free fixed point at the cutoff scale, but we do not start from it in
the UV.  Duality is thus a weaker concept, since we do not know
precisely the Lagrangian at the cutoff.

An example (not necessarily the specific
theory we have discussed in the paper so far) of how small subleading
$1/N$ effects may change the global structure of the flow, and the need
to fine tune at the UV, is shown in fig. 1.

\fig{Global aspects of the flow. Black arrows are the leading N 
contribution. Dashed/White arrows are the subleading N correction.
Line a is the modified flow from the UV fixed point. Line b is the
fine tuned trajectory needed to hit the IR fixed point (we have
neglected the fact that the IR fixed point moves a
bit once $1/N$ corrections are included.}{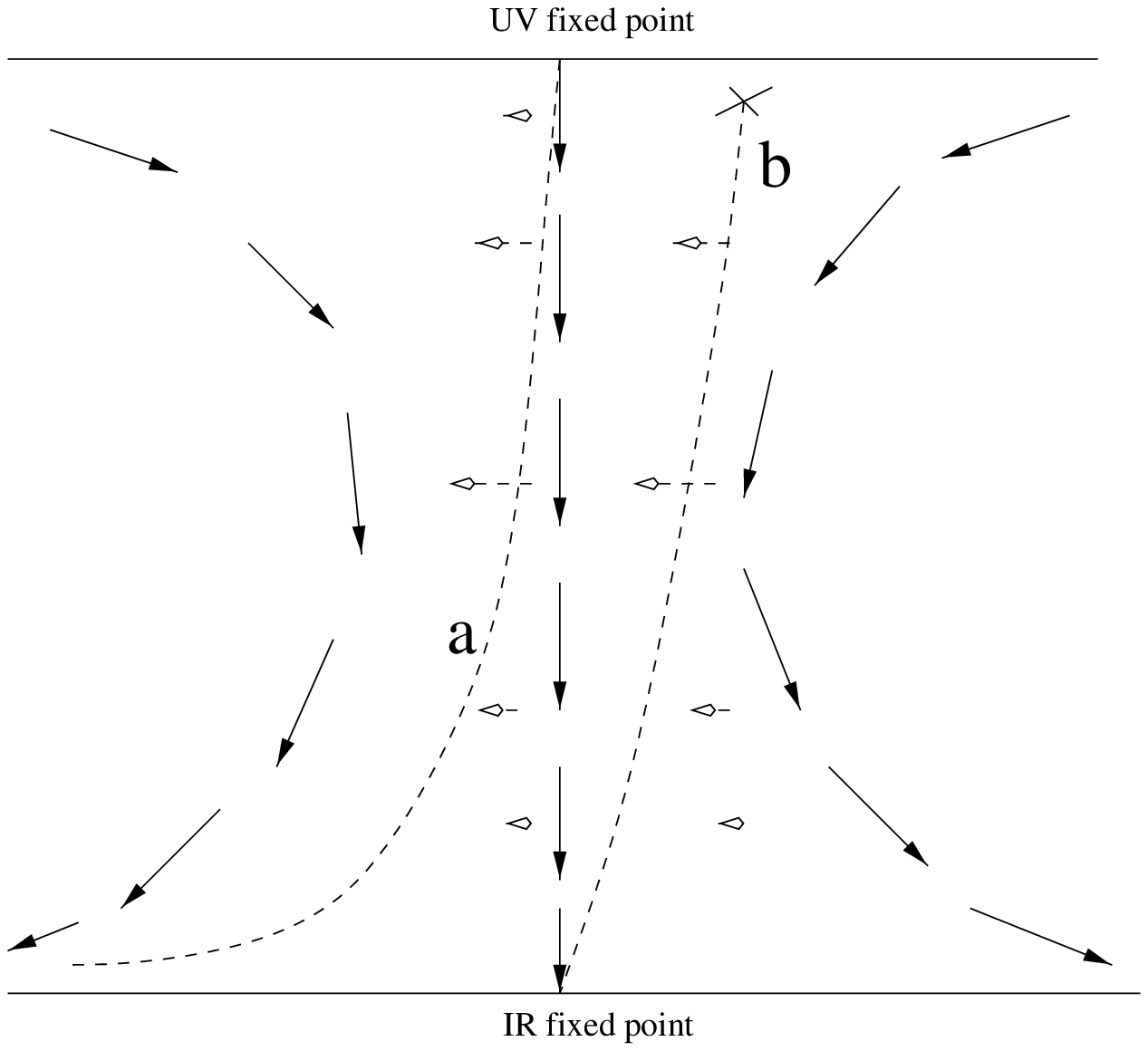}{3.7in}

\medskip
\noindent{\it M-theory considerations}

In the $AdS/CFT$ correspondence the statement that for $N=\infty$ the
RG flow is the same as in $\CN=8$ is mimicked by the fact that the
orbifold of the entire $\CN=8$ solution \itzhaki\ at all scales is
still a solution of the classical equations of motion.

Consider now $1/N$ corrections. These corrections are present at each
value of $U$ (where $U$ is the additional coordinate in the $AdS$,
which contains information about the RG flow). The zeroth order
solution is no longer a solution and we need to correct it. When
correcting it we may either keep the boundary conditions at $U=\infty$
fixed or the behavior at $U=0$ fixed. In the first case we keep the UV
of the theory fixed but then the corrections at $U=0$ may be
significant and the solution there may longer by approaching
$AdS$. Instead we would like to keep the $AdS$ near $U=0$ but we can
do so at the price of maybe changing the $U=\infty$ behavior.

One may ask whether from the supergravity description one can argue
that the field theory becomes a gaussian theory in the UV. It would
seem that the answer is no. The reason is that in the supergravity
solution all that one sees near the boundary of the space-time are
large curvatures \itzhaki. Without independent means of computing at
large curvature, all one can say is that this is consistent with the
field theory becoming weakly coupled in the UV. One may perhaps also
deduce the number of degrees of freedom from black hole entropy
counting, or other dominant effects, but one can not argue that one
knows exactly the Lagrangian of this weakly coupled theory at some
given cutoff.

\subsec{An example of a nonsupersymmetric dual pair}

We need to exhibit two distinct theories which flow in the IR to the
theory of anti-M2 branes near the B-singularity. For example, we may
consider $(\BR^7\times {\bf S}^1)/\BZ_2$ orbifolds of M-theory of
types AB and BB and probe them with anti-M2-branes. At weak coupling
(i.e.  when the radius of ${\bf S}^1$ is small) the M-theory orbifold
of type BB becomes an $\tilde{O2}^+$ plane in IIA, while the orbifold
of type AB becomes an $O2^+$ plane.  Anti-M2 branes become anti-D2
branes in this limit. Naively, one expects the theories of anti-D2
branes probing the $\tilde{O2}^+$ and $O2^+$ planes to be IR dual. As
explained above, this is only literally true for $N=\infty$, and for
finite $N$ one may need to add renormalizable and nonrenormalizable
operators with fine-tuned coefficients in order to preserve
duality. An analogous supersymmetric duality was suggested in
\sethi. The difference is that in the supersymmetric case the theories
have a moduli space of vacua, and to see the duality one needs to go
to a specific place in the moduli space. We have argued above that in
the nonsupersymmetric case the moduli space is lifted at subleading
order in the $1/N$ expansion, so both theories have a unique vacuum
and no tuning of the moduli is necessary.

The theories on anti-D2 branes are of course gauge theories. They are
closely related to $\CN=8$ theories on D2 branes probing the same
backgrounds; in fact, the bosonic fields are identical. To obtain the
spectrum of fermions recall that the field theory on $N$ (anti-)D2
branes near an orientifold 2-plane is obtained by orientifolding the
spectrum of the $\CN=8$ $U(2N)$ theory. In the supersymmetric case the
projection is identical for fermions and bosons, while in the
nonsupersymmetric case the projection for the fermions has an extra
minus sign compared to that for the bosons. It follows that the
spectrum of the gauge theory of $N$ anti-D2 branes near an
$\tilde{O2}^+$ (resp. $O2^+$) orientifold contains gauge bosons and
seven real scalars in the adjoint of $SO(2N+1)$ (resp. $Sp(2N)$) and
eight Majorana fermions in the symmetric tensor representation of
$SO(2N+1)$ (resp. antisymmetric tensor representation of $Sp(2N)$). We
do not know the precise Lagrangian, for reasons explained above. At
leading order in $1/N$ the Lagrangian can be obtained by taking the
corresponding $\CN=8$ Lagrangian describing D2 branes and replacing
fermions in the adjoint by fermions in the appropriate tensor
representation of the gauge group. This Lagrangian is
superrenormalizable. We expect that all terms allowed by symmetries,
including nonrenormalizable ones, would have to be included at
next-to-leading order if one wants to flow to the CFT described by the
``skew-whiffed'' $AdS^4\times\RP7$.

\bigbreak\bigskip\bigskip\centerline{{\bf Acknowledgments}}\nobreak

We would like to thank O. Aharony, S. Kachru, E. Silverstein, and
M. Strassler for useful discussions. The work of MB is supported by
NSF grant PHY-9513835. The work of AK is supported by DOE grant
DE-FG02-90ER40542.

\listrefs 

\end